\documentclass[onecolumn,amssymb, nobibnotes, aps, prb]{revtex4-2}
\usepackage[bookmarks=false]{hyperref}
\usepackage[utf8]{inputenc}
\usepackage[english]{babel}
\usepackage{amssymb}
\usepackage{amsmath}
\usepackage{graphicx}
\usepackage{xcolor}
\usepackage{physics}
\usepackage{float}
\usepackage{textcomp}
\usepackage{gensymb}
\usepackage{hyperref}
\usepackage[normalem]{ulem}
\hypersetup{
     colorlinks   = true,
     citecolor    = blue
}

\renewcommand{\vec}[1]{\boldsymbol{#1}}
\graphicspath{{Figs/}}
\newcommand{\be}{\begin{equation}}
\newcommand{\ee}{\end{equation}}
\newcommand{\bea}{\begin{eqnarray}}
\newcommand{\eea}{\end{eqnarray}}

\def\nn{\nonumber}
\def\lb{\label}
\def\pref{\eqref}

\def\g{\gamma}

\begin{document}

\title{Electrostatic interactions in twisted bilayer graphene} 

\author{Tommaso Cea}
\affiliation{Imdea Nanoscience, Faraday 9, 28015 Madrid, Spain}
\author{Pierre A. Pantale\'on}
\affiliation{Imdea Nanoscience, Faraday 9, 28015 Madrid, Spain}
\author{Niels R. Walet}
\affiliation{Department of Physics and Astronomy, University of Manchester, Manchester, M13 9PY, UK} 
\author{Francisco Guinea}
\affiliation{Imdea Nanoscience, Faraday 9, 28015 Madrid, Spain} 
\affiliation{Donostia International Physics Center, Paseo Manuel de Lardiz\'abal 4, 20018 San Sebasti\'an, Spain}

\date{\today}

\begin{abstract}
The effects of the long range electrostatic interaction in twisted bilayer graphene are described using the Hartree-Fock approximation. The results show a significant dependence of the band widths and shapes on electron filling, and the existence of broken symmetry phases at many densities, either valley/spin polarized, with broken sublattice symmetry, or both. 
\end{abstract}

\maketitle

\maketitle
\section{Introduction}
The discovery of a complex phase diagram in twisted bilayer graphene~\cite{Cao2018,cao_nat18}, which includes superconducting and insulating phases, has led to an extensive study of the role of interactions in the electronic structure of this system (see also\cite{AM20,BDEY20,WMWS20}). The existence of narrow bands can be obtained from a continuum model for a twisted bilayer graphene bilayer~\cite{santos_prl07} (see also~\cite{M10,M11}), as reported in~\cite{Bistritzer_pnas11}. Similar results have been obtained using lattice based tight binding models~\cite{SCVPB10,TMM10}. \textcolor{black}{The two techniques tend to give consistent results.} Twisted bilayer graphene forms a moiré pattern with the lattice period $L = a / [ 2 \sin ( \theta / 2 ) ]$, where $a \approx 2.46 \text{\AA}$ is the lattice constant of graphene, and $\theta$ is the twist angle expressed in radians. 
The reported narrow bands occur for twist angles $\theta \approx 1^\circ$, and thus for a moiré lattice vector of $L  \approx a / \theta \approx 15\,\text{nm}$. Typical widths of the bands nearest to the Fermi energy range from $W \sim 1 - 5$ meV. The central bands accommodate an electron density of order $-4 \times  L^{-2} \lesssim 4 \times \rho \lesssim \sim L^{-2}$, where $L^{-2} \sim  10^{12} \mathrm{ cm}^{-2}$. The factor $4$ arises from the spin and valley degeneracy of each band. The fact that $L \gg a$ allows us to classify the strength of interactions in terms of their dependence on $a / L \sim 10^{-2}$, or, alternatively, in terms of the number of carbon atoms in the unit cell, $N \sim ( L / a )^2 \sim 10^4$. 
n inhomogeneous charge distribution on the scale of $L$ leads to electrostatic energies of the order $e^2 / ( \epsilon L ) \sim 10 - 50$ meV, where $\epsilon \sim 4 - 10$ is the dielectric constant of the environment~\cite{Guinea_pnas18}. This energy scales as $L^{-1} \sim 1/\sqrt{N}$. 
The existence of atomic orbitals modifies the electron-electron repulsion at length scales comparable to the lattice constant, $a$. This effect is best described by including an intra-atomic Hubbard term,  $U_H \lesssim e^2 / a \sim 2 - 6$ eV, and other inter-atomic interactions. Two electrons in the central bands have a probability $\sim 1 / N^2$ of being on the same atom. Hence, the contribution of a Hubbard term to the energy of the central bands per moiré unit cell is $\sim U_H \times N \times 1 / N^2 \sim U_H / N$. 

A similar analysis can be carried out for the electron-electron interactions mediated by phonons. Longitudinal acoustic phonons are coupled to electrons by the deformation potential, $D \approx 20 - 30$ eV. In Fourier space, this coupling translates into an effective electron-electron interaction $V_{e-e}^{ph_l} \sim D^2 / ( \lambda + 2 \mu )$, where $\lambda$ and $\mu$ are the Lam\'e coefficients which describe the elastic properties of a graphene layer. By averaging this coupling over the moir\'e scale, we obtain a coupling  $V_{e-e}^{ph_l} \sim D^2 / ( \lambda + 2 \mu ) \times L^{-2} \sim 1/N$. Transverse and optical phonons do not induce local changes in the chemical potential, and couple to electrons only through a modification of the hopping parameters, which can be described as an effective gauge field. The dimensionless parameter which describes this coupling is $\beta = ( a / t ) \partial t / \partial a \sim 3$, where $t$ is the nearest neighbor hopping. Transverse acoustic phonons lead to an effective interaction in Fourier space $\sim ( t^2 \beta^2 ) / \mu$, which translates into an effective interaction averaged over the moiré unit cell into $V_{e-e}^{ph_{tr}} \approx (t^2 \beta^2 )/ ( a^2 \mu ) \times L^{- 2} \sim 1/N$. For optical phonons, the effective electron-electron coupling defined in Fourier space is $ ( t / a )^2 \beta^2 ( \rho a^2 \omega_{ph}^2 )^{-1}$, where $\rho$ is the mass density, and $\omega_{ph}$ is the frequency of the relevant phonon. In real space, averaging over the moir\'e unit cell, the previous quantity defines an energy scale $V_{e-e}^{ph_{opt}} \sim ( t^2 \beta^2 )/ ( \rho a^2 \omega_{ph}^2 ) \times L^{-2} \sim 1/N$. A summary of the different interactions is shown in Table~\ref{tbl:interactions}.

\begin{table}
\renewcommand{\arraystretch}{2}
\begin{tabular}{||c|c|c||}
\hline\hline
Interaction & \, \, \, \, Energy scale \, \, \, \, & Dependence on the number of atoms per unit cell \\ \hline \hline
Long range electrostatic interaction & $\displaystyle \frac{e^2}{\epsilon L}$ & $\displaystyle\frac{1}{\sqrt{N}}$ \\ \hline
Intra-atomic interaction &$U_H \times \frac{ a^2}{L^2}$ &$\displaystyle\frac{1}{N}$ \\ \hline
Longitudinal acoustic phonons &$\displaystyle \frac{D^2}{( \lambda + 2 \mu ) L^2}$ &$\displaystyle \frac{1}{N}$ \\ \hline
Transverse acoustic phonons &$\displaystyle \frac{t^2 \beta^2}{( a^2 \mu)} \times \frac{a^2}{L^2}$ &$\displaystyle \frac{1}{N}$ \\ \hline
Optical phonons & $\displaystyle \frac{t^2 \beta^2}{\rho a^2 \omega_{ph}^2}  \times \frac{1}{L^2}$ & $\displaystyle \frac{1}{N}$ \\
\hline\hline
\end{tabular}
\caption{Interactions (left), typical energies averaged over the moir\'e unit cell (center), and dependence on the number of atoms in the unit cell (right).}
\label{tbl:interactions}
\end{table}

The qualitative analysis presented here shows that the leading electron-electron interaction in twisted bilayer graphene at small angles, $\theta \sim 1^\circ$, where the number of atoms in the unit cell is large, $N \sim 10^4 \gg 1$, is the long range electrostatic interaction. In most crystalline solids, the leading effects of this interaction can be captured by the Hartree-Fock approximation. Moreover, twisted bilayer graphene at sufficiently high temperatures shows a metallic behavior, with features consistent with the existence of a Fermi surface. This metallic phase can be expected to be amenable to a mean field, Hartree-Fock, description. 

In the following, we describe the main features obtained so far in Hartree-Fock analyses of twisted bilayer graphene. \textcolor{black}{This approximation is standard in quantum mechanics, and it gives useful insights in theoretical condensed matter physics.} We will focus mostly on studies carried out using the continuum model for the electronic structure of twisted bilayer graphene \cite{Bistritzer_pnas11}, although lattice-based tight-binding models have also been considered~\cite{SCVPB10,TMM10}. Using the continuum approximation, the problem is defined in terms of three quantities with dimensions of energy,  which describe the non interacting band, $v_F / L , g_1$ and $g_2$. The parameter $v_F$ is the Fermi velocity, and $g_1$ and $g_2$ are averages of the interlayer hopping, see below. Finally the interactions are described by the parameter $e^2/ \epsilon$, where $e$ is the electric charge, and $\epsilon$ is the dielectric constant of the environment.

This brief review is focused on recent results on the electronic structure of twisted bilayer graphene, obtained within the Hartree-Fock theory. It is worth mentioning that the Hartree-Fock approximation has been extensively used for the analysis of interactions in two dimensional materials~\cite{KUPGN12}, including the energy splittings of flat Landau levels\cite{G11}. This paper does not attempt to cover in full the already extensive literature devoted to this topic. We leave out analyses beyond mean field, such as RPA calculations of the low energy excitations~\cite{LL19,KIVF19,WS20,STSVA20,KBVZ20,KXM20,KZB20,KCBZV21,CG21}. The main conclusions, which description a number of novel features almost unique to twisted bilayer graphene, are consistent with most studies carried out to date. 

The next section discusses the main features of the model. Then, we describe the implementation of the Hartree-Fock approximation to the study of long range interactions in the model. The following section presents the dependence of the electronic bands on band filling and twist angle. We discuss next possible broken symmetry phases. Finally, we highlight a number of novel features induced by long range electrostatic interactions, treated within the Hartree-Fock approximation, 

\section{The Hartree-Fock approximation within the continuum model of TBG}
\begin{figure}[ht!]
\includegraphics[width=2.5in]{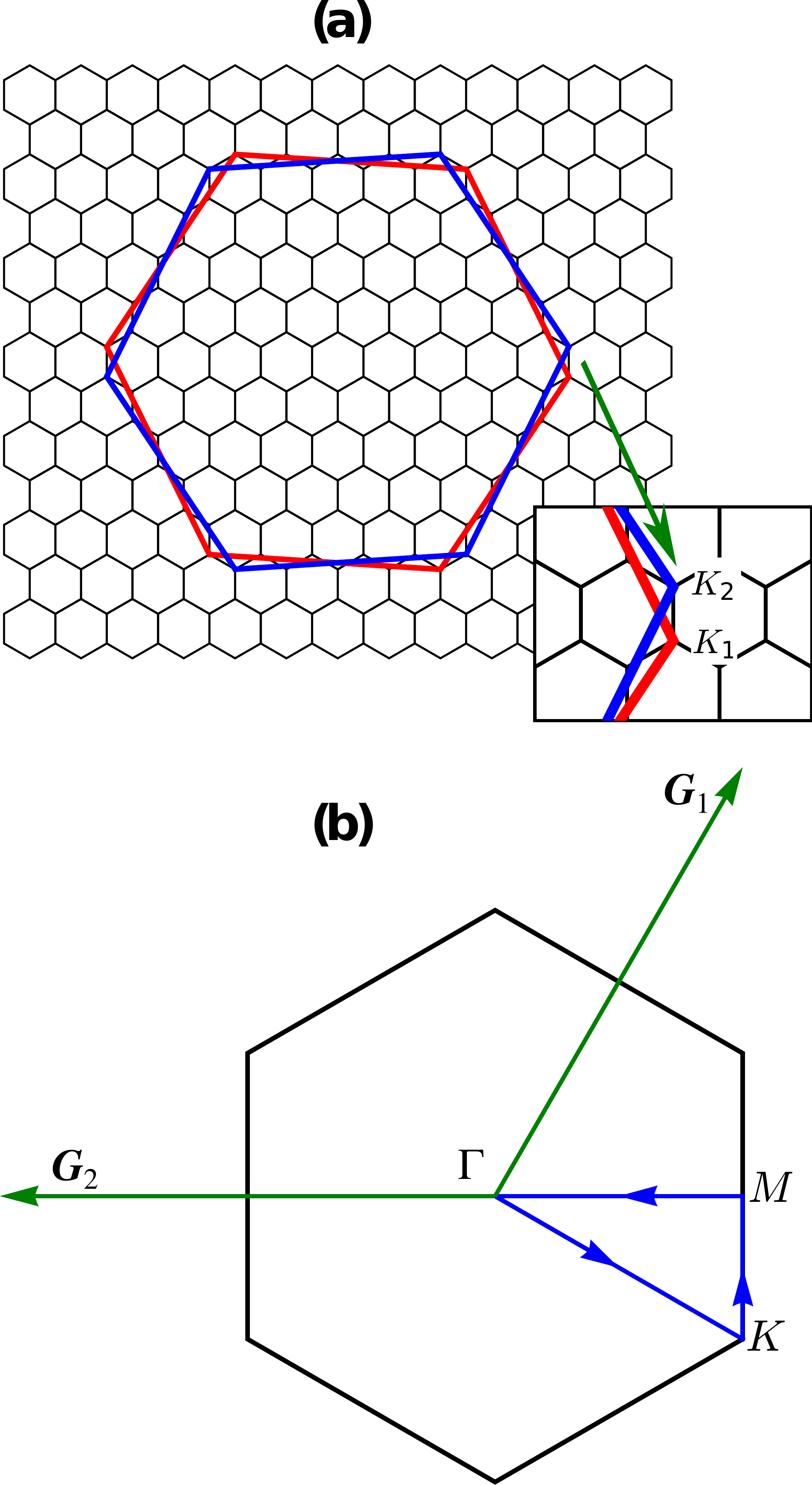}
\caption{
(a) Folding of the BZs of each of the monolayers in bilayer graphene.
The BZ of layer 1 (indicated by the red hexagon) is rotated by $-\theta/2$, while that of the layer 2 (the blue hexagon) is rotated by $\theta/2$.
The small black hexagons represent the unit cells forming the reciprocal moir\'e lattice. In the inset we show $K_{1,2}$ are the Dirac points of the twisted monolayers, which identify the corners of the mini-BZ~\cite{koshino_prx18}. (b) mini-BZ. $\vec{G}_{1,2}$ are the two reciprocal lattice vectors. The blue line shows the high symmetry circuit in the mini-BZ used to compute the bands shown in the following.}
\label{BZ}
\end{figure}

Rotating two layers of graphene by a relative small angle, $\theta$, gives rise to a moir\'e pattern. The period of the moir\'e, $L=\frac{a}{2\sin(\theta/2)}$, dramatically increases by reducing $\theta$. We describe the TBG within the low energy continuum model considered in Refs.~\cite{santos_prl07,Bistritzer_pnas11,santos_prb12,koshino_prx18}, which is meaningful for sufficiently small angles, so that an approximately commensurate structure can be defined for any twist. The moir\'e mini-BZ, resulting from the folding of the two BZs of each monolayer (see Fig.\ref{BZ}(a)), is generated by the two reciprocal lattice vectors:
\begin{equation}
\vec{G}_1= \frac{2\pi}{L}\left(\frac{1}{\sqrt{3}},1 \right)\text{ and } \vec{G}_2=\frac{4\pi}{L}\left(-\frac{1}{\sqrt{3}},0 \right),
\end{equation}
 shown in green in Fig.~\ref{BZ}(b). 
Let $K_\xi=\xi4\pi(1,0)/3a$ be the two Dirac points of the unrotated monolayer graphene, with $\xi=\pm1$. For small twists, the coupling between the $K_+$ and $K_-$ valleys of the two monolayers can be safely neglected, as the interlayer hopping has a long wavelength modulation. The fermionic field operators of the TBG are  4-component Nambu spinors:
\bea
\Psi_{\xi\sigma}=\left(\psi_{\xi\sigma}^{A_1},\psi_{\xi\sigma}^{B_1},\psi_{\xi\sigma}^{A_2},\psi_{\xi\sigma}^{B_2}\right)^T,
\eea 
where $A,B$ and the subscripts $1,2$ denote the sub-lattice and layer indices, respectively, and $\sigma$ is the spin index. We introduce a relative twist $\theta$ between the two monolayers  by rotating the layer $1$ by $-\theta/2$ and the layer $2$ by $\theta/2$.
Without loss of generality, we assume that in the aligned configuration, at $\theta=0$, the two layers are $AA$-stacked.
In the continuum limit, the effective Hamiltonian of the TBG in a volume $\Omega$ can be generally written as
~\cite{santos_prl07,Bistritzer_pnas11,santos_prb12,koshino_prx18}:
\bea\lb{HTBG}
\hat{H}_{TBG}=\sum_{\xi\sigma}\int_\Omega\,d^2\vec{r}\Psi_{\xi\sigma}^\dagger(\vec{r})
\begin{pmatrix}
H_{\xi1}&U_\xi(\vec{r})\\U_\xi^\dagger(\vec{r})&H_{\xi2}
\end{pmatrix}
\Psi_{\xi\sigma}(\vec{r}),
\eea
where
\bea
H_{\xi l}=\xi \hbar v_F \left(-i\vec{\nabla}-\xi K_l\right)\cdot
\vec{\tau}^\xi_{\theta_l}
\eea
is the Dirac Hamiltonian for the $\xi$ valley of layer $l$, $v_F=\sqrt{3}ta/(2\hbar)$ is the Fermi velocity, $t$ is the hopping amplitude between localized $p_z$ orbitals at nearest neighbors carbon atoms, $\theta_{1,2}=\mp\theta/2$, $\vec{\tau}^\xi_{\theta_l}=e^{i\xi\tau_z\theta_l/2}\left(\tau_x,\xi\tau_y\right)e^{-i\xi\tau_z\theta_l/2}$, $\tau_i$ are the Pauli matrices, and $K_l=4\pi\left(\cos\theta_l,\sin\theta_l\right)/(3a)$ are the Dirac points of the two twisted monolayers corresponding to the $\xi=+$ valley, which identify the corners of the mini-BZ shown in Fig.~\ref{BZ}(a). $U_\xi(\vec{r})$ is the inter layer potential, which is a periodic function in the moir\'e unit cell. In the limit of small angles, its leading harmonic expansion is determined by only three reciprocal lattice vectors ~\cite{santos_prl07}:
$U_\xi(\vec{r})=U_\xi(0)+U_\xi\left(-\vec{G}_1\right)e^{-i\xi\vec{G}_1\cdot\vec{r}}+
U_\xi\left(-\vec{G}_1-\vec{G}_2\right)e^{-i\xi\left(\vec{G}_1+\vec{G}_2\right)\cdot\vec{r}}$, where the amplitudes $U_\xi\left(\vec{G}\right)$ are given by:
\bea
U_\xi(0)&=&\begin{pmatrix}g_1&g_2\\g_2&g_1\end{pmatrix},\nn\\
U_\xi\left(-\vec{G}_1\right)&=&\begin{pmatrix}g_1&g_2e^{-2i\xi\pi/3}\\g_2e^{2i\xi\pi/3}&g_1\end{pmatrix},\\
U_\xi\left(-\vec{G}_1-\vec{G}_2\right)&=&\begin{pmatrix}g_1&g_2e^{2i\xi\pi/3}\\g_2e^{-2i\xi\pi/3}&g_1\end{pmatrix}.\nn
\eea
In the following we adopt the parametrization of the TBG given in the Ref.~\cite{koshino_prx18}: $\hbar v_F/a=2.1354$eV,
$g_1=0.0797$eV and $g_2=0.0975$eV. The difference between $g_1$ and $g_2$, as described in~\cite{koshino_prx18}, accounts for the inhomogeneous interlayer distance, which is minimum in the $AB/BA$ regions and maximum in the $AA$ ones, or it can be seen as a model of a more complete treatment of lattice relaxation~\cite{guinea_prb19}. In that latter reference it is also explained that one may want to use the full dispersion of the graphene bands, and it is shown that in general the Fermi surface shifts as a function of twist angle. If we focus, e.g., on the $\xi=+$ valley, then the Hamiltonian of Eq.~\pref{HTBG} hybridizes states of layer $1$ with momentum $\vec{k}$ close to the Dirac point with the states of layer $2$ with momenta $\vec{k},\vec{k}+\vec{G}_1,\vec{k}+\vec{G}_1+\vec{G}_2$.

\begin{figure}
\includegraphics[width=4.5in]{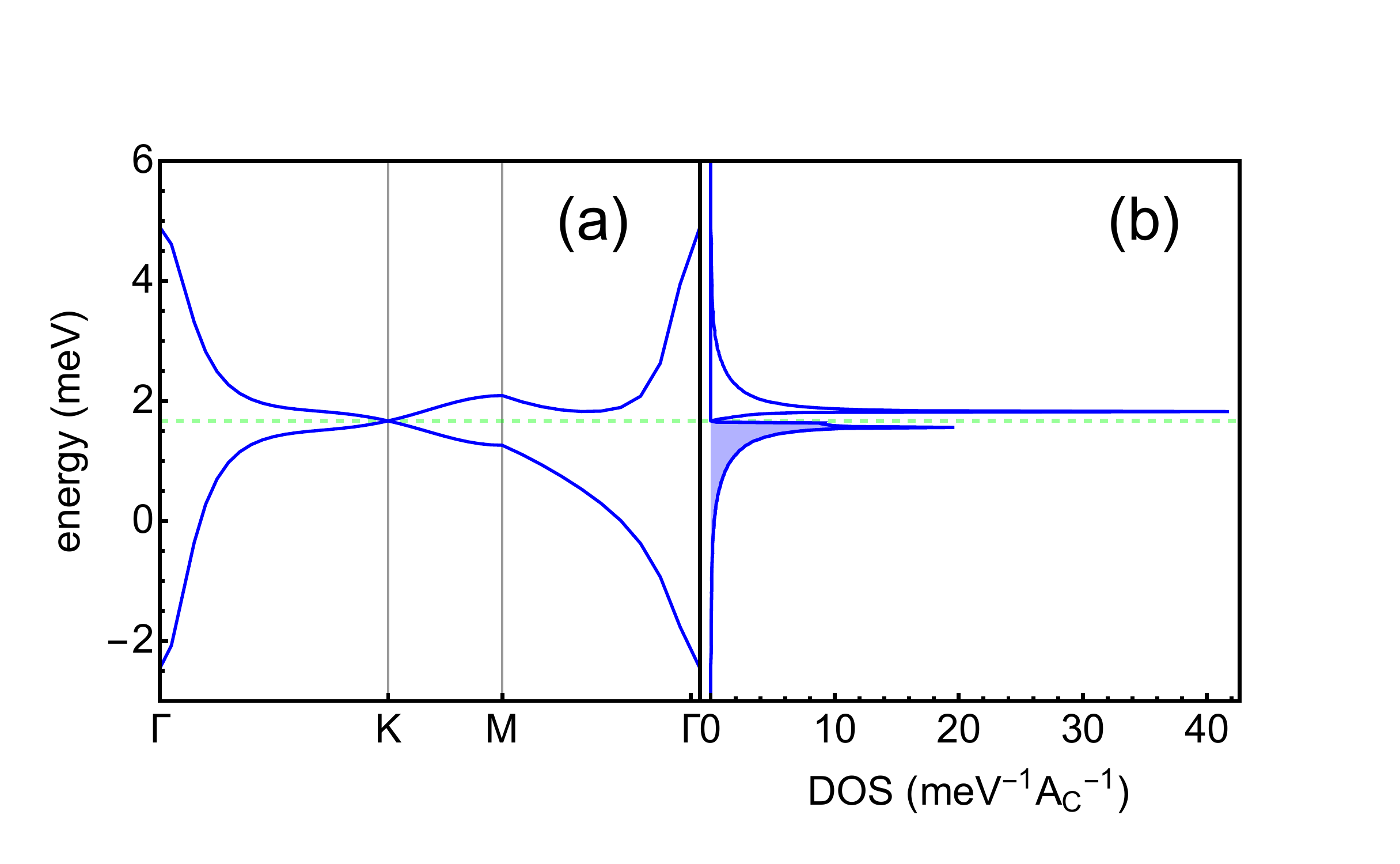}
\caption{
(a) mini-bands of the non-interacting TBG at CN, obtained for the twist angle $\theta=1.05^\circ$ and computed along the high symmetry circuit of the BZ denoted by the blue arrows in Fig.~\ref{BZ}(b).
The green dashed line identifies the Fermi energy.
(b): DOS per moir\'e unit cell, normalized to 8.
}
\label{bands_koshino}
\end{figure}

In the absence of interactions, the band structure and the DOS per moir\'e unit cell of the mini-bands at charge neutrality (CN) are shown in Fig.~\ref{bands_koshino}, for $\theta=1.05^\circ$. The green dashed line identifies the Fermi energy. The bands are computed along the high symmetry circuit of the BZ denoted by the blue arrows in Fig.~\ref{BZ}(b). Here we are showing only one valley, the opposite one being related by the time-reversal symmetry, upon inverting $\vec{k}$ to $-\vec{k}$. $A_C=L^2\sqrt{3}/2$ is the area of the moir\'e unit cell and the DOS is normalized to 8, accounting for two bands and four spin/valley flavors. As deeply studied in the past literature~\cite{santos_prl07,Bistritzer_pnas11}, these bands are characterized by a very narrow bandwidth, $\lesssim 5$ meV,
and by an almost vanishing Fermi velocity as compared to that of monolayer graphene, thus pinning the DOS at the two van Hove singularities in Fig.~\ref{bands_koshino}(b). 

Next we introduce the Coulomb interaction, as described by the Hamiltonian
\bea\lb{H_C}
\hat{H}_C=\frac{1}{2}\int_\Omega\,d^2\vec{r}d^2\vec{r}'\delta\hat{\rho}(\vec{r})v_C(\vec{r}-\vec{r}')\delta\hat{\rho}(\vec{r}'),
\eea
where $\delta\hat{\rho}(\vec{r})\equiv \hat{\rho}(\vec{r})-\rho_{CN}(\vec{r})$ is the quantum operator associated to the density fluctuations, $\hat{\rho}(\vec{r})=\sum_{ \mu}\Psi^\dagger_\mu(\vec{r})\Psi_\mu(\vec{r})$ is the density operator, $\mu=(\xi,\sigma)$ are the generalized valley and spin index, $\rho_{CN}(\vec{r})$ is the average density corresponding to the non-interacting TBG at CN, and $v_C(\vec{r})$ is the Coulomb potential. In the following we assume that the Coulomb interaction is actually screened by a double metallic gate, and thus described by the Fourier transform
\begin{equation} v_C(\vec{q})\equiv\int\,d^2\vec{r}v_C(\vec{r})e^{-i\vec{q}\cdot\vec{r}}=\frac{2\pi e^2}{\epsilon |\vec{q}|}\tanh\left(|\vec{q}|d\right),\end{equation} where $e$ is the electron charge, $\epsilon$ the dielectric constant of the environment and $d$ the distance of the sample from the gate.

The calculations to be discussed have been carried out in momentum space, using the continuum model. The Fourier transform of the density operator is
\begin{align}
    \hat{\rho}_{\vec{q}} &= e^{i \vec{q} \vec{r}} {\cal I}_s {\cal I}_\tau {\cal I}_\sigma {\cal I}_{l}
    \label{densq}
\end{align}
where $\vec{r}$ is the position operator in the continuum model, and ${\cal I}_i$ is a unit $2 \times 2$ matrix in the spin, valley, sublattice, and layer degrees of freedom. 
The mean field analysis described below can be expressed in terms of expectation values of the operator in Eq.~(\ref{densq}). Of particular relevance are the expectation values (or form factors) of the type $\langle \vec{k} , m | \hat{\rho}_{\vec{q}} | \vec{k} + \vec{q} , n \rangle$, where the indices $m$ and $n$ stand for the central bands. Results in the following sections have been calculated using the parameters $\epsilon=10$ and $d=40~\text{nm}$, which are realistic values for most experiments. At mean-field level, the Hamiltonian $\hat{H}_C$ is replaced by
\bea\label{HCMF}
\hat{H}_C\rightarrow \hat{H}^{MF}_C=\hat{H}_H+\hat{H}_F+E_0,
\eea 
where
\begin{subequations}\lb{HFterms}
\bea\lb{Hartree}
\hat{H}_H=\sum_{i\mu}\int_\Omega\,d^2\vec{r}\psi^{i,\dagger}_\mu(\vec{r})\psi^{i}_\mu(\vec{r}) V_H(\vec{r})
\eea
is the Hartree Hamiltonian and $V_H(\vec{r})=\int_\Omega\,d^2\vec{r}'v_C(\vec{r}-\vec{r}')\left\langle\delta\hat{r}(\vec{r}')\right\rangle$ is the local Hartree potential;
\bea\lb{Fock}
\hat{H}_F=\sum_{ij\mu}\int_\Omega\,d^2\vec{r}\,d^2\vec{r}'\psi^{i,\dagger}_\mu(\vec{r})V^{ij}_{F,\mu}(\vec{r},\vec{r}')\psi^{j}_\mu(\vec{r}')
\eea
is the Fock Hamiltonian and $V^{ij}_{F,\mu}(\vec{r},\vec{r}')=-\left\langle \psi^{j,\dagger}_\mu(\vec{r}')   \psi^{i}_\mu(\vec{r})\right\rangle v_C(\vec{r}-\vec{r}') $ is the non-local Fock potential.

The Hartree potential is absent in single layer graphene, and in aligned bilayers, because the charge distribution is homogeneous, and compensated by the external gate. On the other hand, the exchange (Fock) potential is present in monolayer and bilayer graphene. Its effect is by no means negligible, as it causes the renormalization of the Fermi velocity~\cite{NGPNG09}. In very clean graphene samples, and at charge neutrality, the Fermi velocity diverges logarithmically as function of a cut-off defined by the ratio between the bandwidth and a low energy energy cutoff set by the Fermi energy, or by the fluctuations in the local charge due to disorder. Part of the effects of this exchange potential is already included in the effective parameters which describe the bands of graphene. When calculating the exchange potential in twisted bilayer graphene this renormalization needs to be taken into account. Different  procedures have been proposed~\cite{XM20,bultinck_prx20}, such as the subtraction of the exchange potential in a monolayer, or the subtraction of the exchange potential due to the non interacting central bands in twisted bilayer graphene. The exchange potential per graphene unit cell in one monolayer scales as $\Sigma_x \propto e^2 \Lambda$, where $\Lambda \sim 1/a$ is a momentum cut-off comparable to the lattice unit cell. Hence, the effect of the central bands in twisted bilayer graphene can be expected to be of order $\sim e^2 / L$ per moiré unit cell of side $L$. In the following, we will consider that the contribution of the central bands to the exchange energy in monolayer graphene can be neglected, and that no subtraction is required. We have,
\bea
E_0&=&-\frac{1}{2}\left[\left\langle   \hat{H}_H+ \hat{H}_F \right\rangle\
+\int_\Omega\,d^2\vec{r}\rho_{CN}(\vec{r})V_H(\vec{r})\right]
\eea
as the zero point energy, which is required to avoid double counting of the total energy at mean-field level.
\end{subequations}
The mean-field Hamiltonian for the interacting TBG is then:
\bea\label{HMF}
\hat{H}^{MF}=\hat{H}_{TBG}+\hat{H}_C^{MF},
\eea
which we diagonalize self-consistently, by computing the quantum averages of the Eqs. \pref{HFterms} by means of $\hat{H}^{MF}$
and iterating until convergence. It is worth noting that this procedure is equivalent to minimize the GS energy of $\hat{H}^{MF}$.
In order to diagonalize $\hat{H}^{MF}$, we exploit the Bloch's theorem, by expressing the eigenfunctions in the basis of Bloch's plane waves defined on the moir\'e:
\bea\lb{eigenstates}
\ket{\vec{k},\alpha,\mu}=\sum_{\vec{G}i}\phi_{\vec{k}+\vec{G},\alpha,\mu,i}\ket{\vec{k}+\vec{G},\mu,i},
\eea 
where $\vec{k}\in$mBZ, the $\vec{G}$'s are reciprocal lattice vectors, $\alpha$ is the band index and $\phi_{\vec{k}+\vec{G},\alpha,\mu,i}$ are numerical eigenvectors normalized according to:
$\sum_{i\vec{G}}\phi^*_{\vec{k}+\vec{G},\alpha,\mu,i}\phi_{\vec{k}+\vec{G},\alpha',\mu,i}=\delta_{\alpha \alpha'}$.
Upon using the Eq.~\pref{eigenstates} to evaluate the Hartree and Fock potentials, the matrix elements of the Eqs. \pref{HFterms} can be written in the Bloch's basis as: 
\begin{subequations}\lb{matrixelements}
\bea\lb{Hartree_mat_el}
\bra{\vec{k}+\vec{G},\mu,i}\hat{H}_H\ket{\vec{k}+\vec{G}',\mu',i'}=
\delta_{ii'}\delta_{\mu\mu'}\frac{v_C(\vec{G}-\vec{G}')}{\Omega}
\times\nn\\
\times\sum_{\vec{k}'\vec{G}''}\sum_{\alpha \mu'' i''}
\phi_{\vec{k}'+\vec{G}''+\vec{G},\alpha,\mu'',i''}
\phi^*_{\vec{k}'+\vec{G}''+\vec{G}',\alpha,\mu'',i''}\equiv\delta_{ii'}\delta_{\mu\mu'} V_H\left(\vec{G}-\vec{G}'\right),
\eea
where the sum over the band index, $\alpha$, runs over all the occupied states counted from CN, and $V_H\left(\vec{G}\right)$ is noting but the Fourier transform of the Hartree potential, $V_H\left(\vec{r}\right)$, evaluated in $\vec{G}$.
\bea\lb{Fock_mat_el}
\bra{\vec{k}+\vec{G},\mu,i}\hat{H}_F\ket{\vec{k}+\vec{G}',\mu',i'}&=&\nn\\
-\delta_{\mu\mu'}\sum_{\vec{k}'\vec{G}''\alpha}\frac{v_C(\vec{k}-\vec{k}'-\vec{G}'')}{\Omega}
\phi_{\vec{k}'+\vec{G}''+\vec{G},\alpha,\mu,i}\phi^*_{\vec{k}'+\vec{G}''+\vec{G}',\alpha,\mu',i'}&\equiv&
\delta_{\mu\mu'}
 \Sigma_F^\mu ( \vec{k} + \vec{G} , i ; \vec{k} + \vec{G}' , i' ),
\eea
\end{subequations}
where $\alpha$ runs over all the occupied states above a given threshold. In the present context, we set this threshold to the lowest energy of the mini-bands in the middle of the spectrum, meaning that we are neglecting the contribution of the bulk bands. However, including other bands might affect quantitatively the results. Because the Eqs. \pref{matrixelements} express the matrix elements in terms of the energy levels and of the corresponding eigenfunctions, $\phi$, they completely define the self-consistent problem.

The main contributions of the long-range interaction are expected to come from small momenta. Therefore, we only consider the matrix elements of the Hartree potential, Eq.~\pref{Hartree_mat_el}, with $\vec{G}-\vec{G}'$ belonging to the first star of reciprocal lattice vectors: $\pm\vec{G}_1,\pm\vec{G}_2,\pm\left(\vec{G}_1+\vec{G}_2\right)$. Concerning the matrix elements of the Fock potential, Eq.~\pref{Fock_mat_el}, for each external momentum $\vec{k}$ we truncate the sum over $\vec{k}'$ and $\vec{G}''$ so that: $\vec{k}-\vec{k}'-\vec{G}''$ belongs to the BZ. We checked that including larger momenta affects negligibly the results.

Finally, the energy of the GS, as following from the Eqs.~\pref{HCMF}-\pref{HMF}, is given by: 
\bea
E_{GS}=\sum_{\vec{k}\alpha\mu}\varepsilon_{\vec{k}\alpha\mu}+E_0,
\eea
where $\varepsilon_{\vec{k}\alpha\mu}$ are the single-particle energies and the sum over $\alpha$ runs over all the occupied states.

\begin{figure*}
\centering
\includegraphics[scale=0.20]{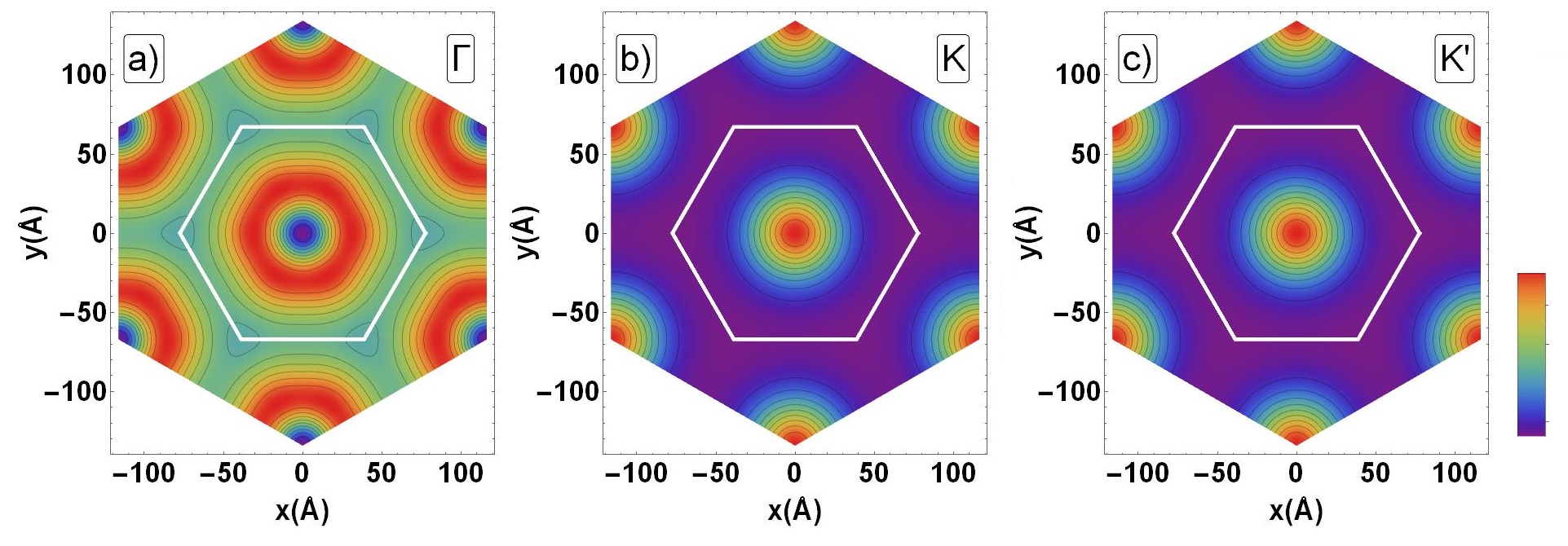}
\caption{Charge density at the high symmetry points within the mBZ for the lower band in Fig.~\ref{bands_koshino}. White hexagon is the real space unit cell.}
\label{Density}
\end{figure*}

\section{Evolution of the band structure as function of filling}

\begin{figure}
\includegraphics[width=3.5 in]{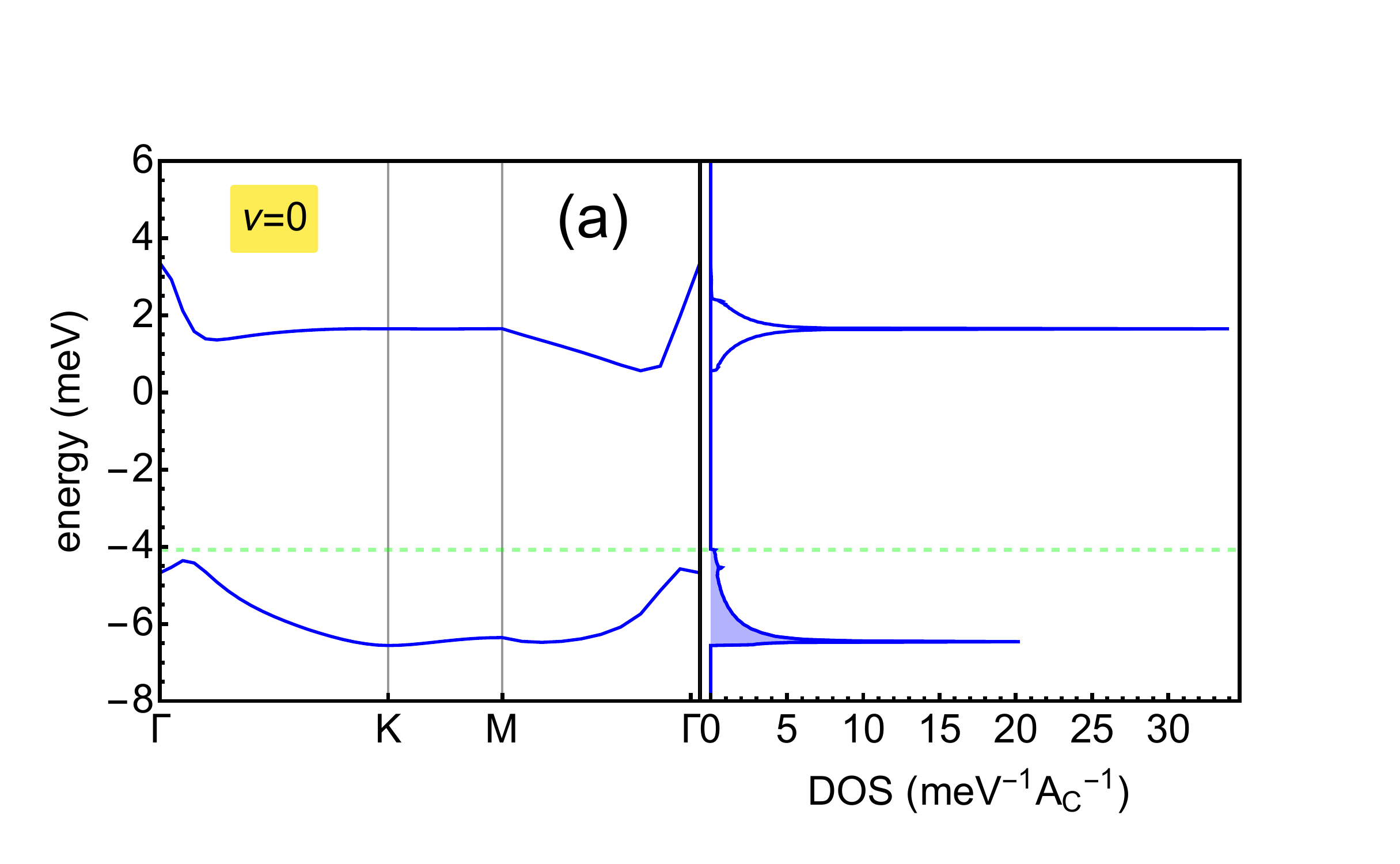}
\includegraphics[width=3.5 in]{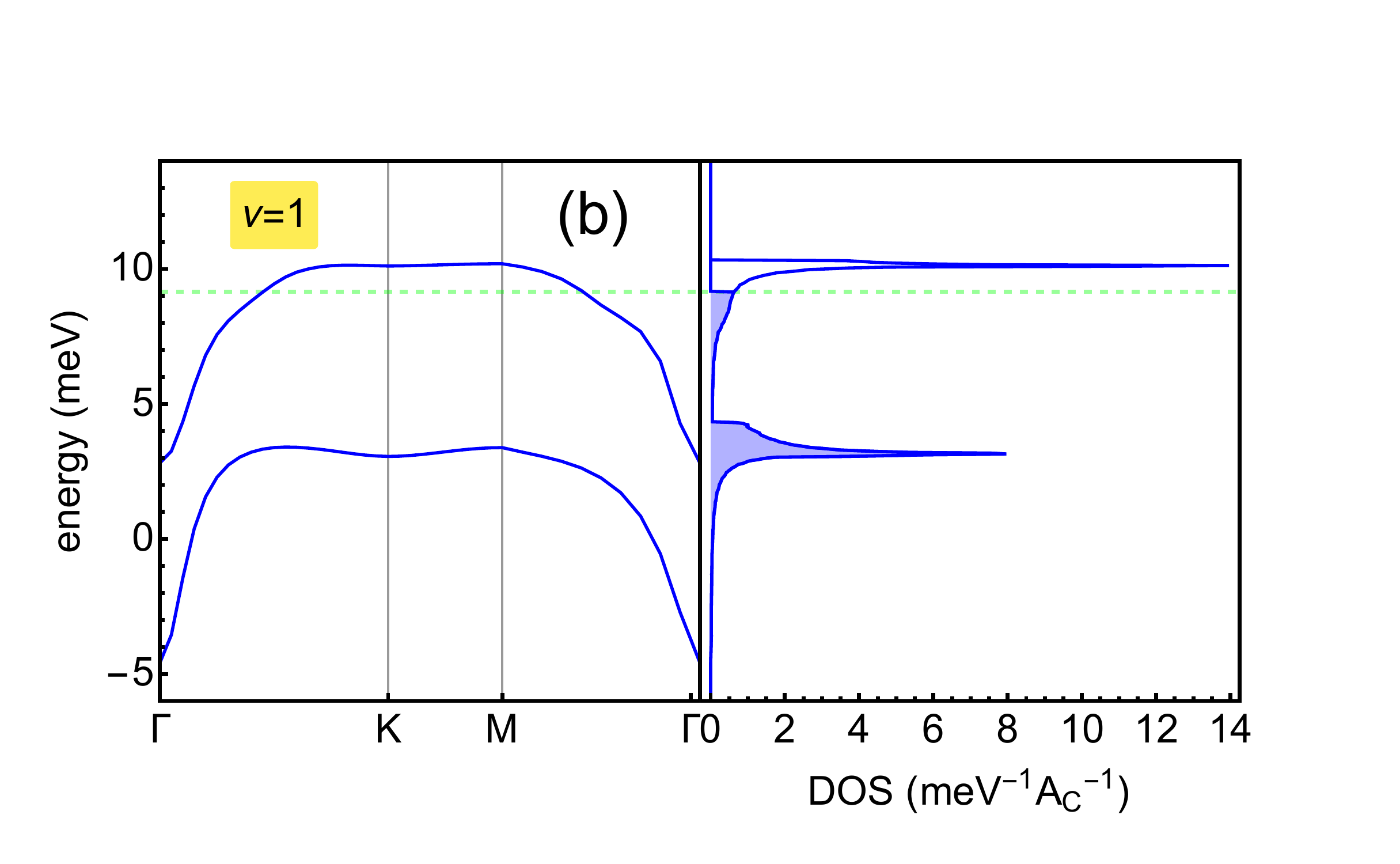}
\includegraphics[width=3.5 in]{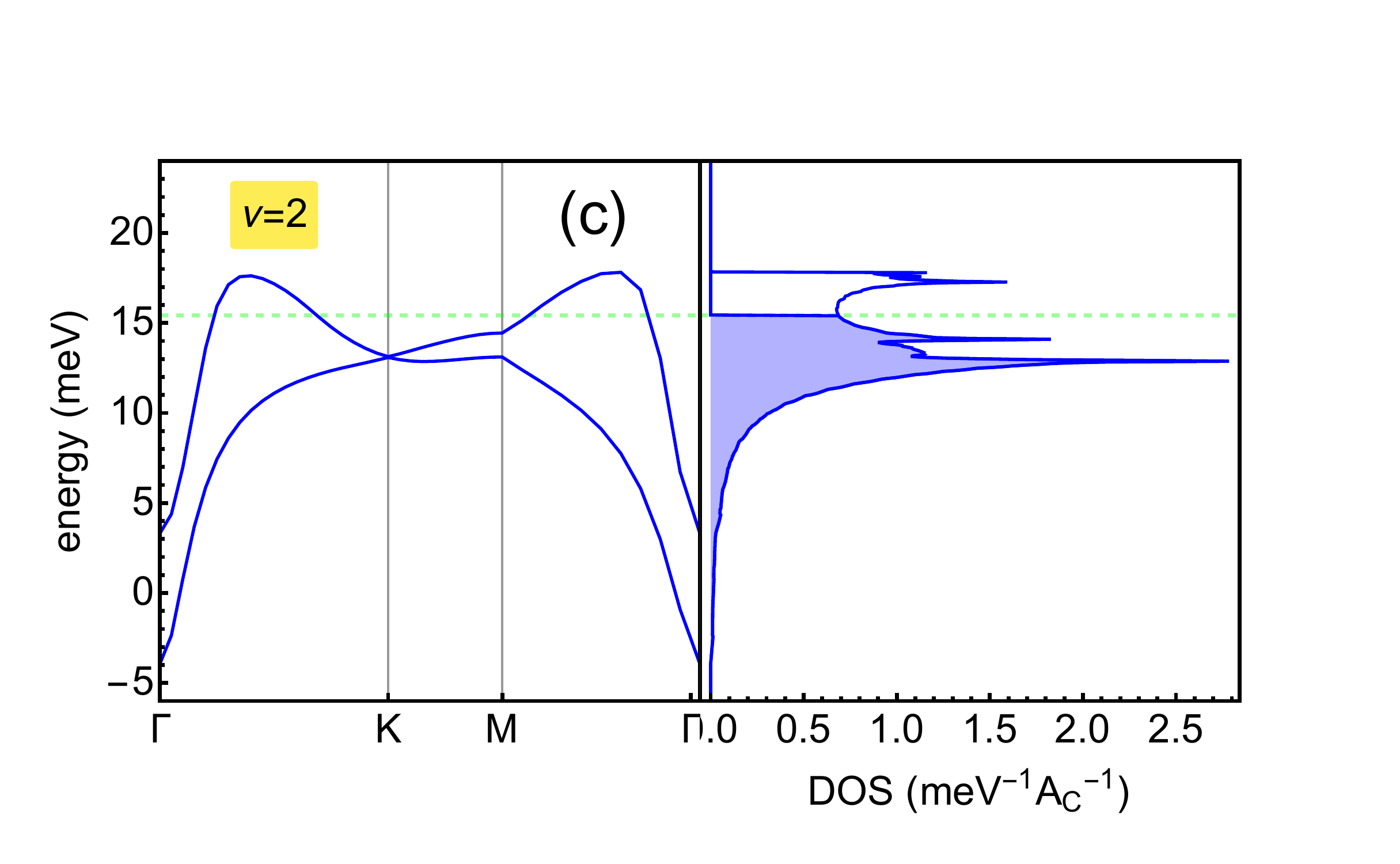}
\caption{Band structure and DOS corresponding to the non-polarized GS, obtained within the Hartree-Fock approximation at the twist angle $\theta=1.05^\circ$ and filling: $\nu=0(a),1(b),2(c)$. The green dashed line identifies the Fermi energy.}
\label{bands_np}
\end{figure}

Figure \ref{bands_np} shows the band structure and DOS of the mini-bands of the interacting TBG at $\theta=1.05^\circ$, for the non-polarized GS at positive filling of the conduction bands: $\nu=0,1,2$, corresponding to CN, one and two electrons per moir\'e unit cell, respectively. The green dashed line identifies the Fermi energy. Each band is four-fold degenerate, so that the spin/valley flavors are equally occupied. At $\nu=0$, only the lower band is filled, the $\mathcal{C}_2$ symmetry is broken and the Fermi surface (FS) is fully gapped. For this choice of parameters, the width of the gap is $\sim 5$meV, comparable to the overall bandwidth. At $\nu=1$, one quarter of the upper band is filled, the $\mathcal{C}_2$ symmetry is still broken, but the FS exists. At $\nu=2$, the $\mathcal{C}_2$ symmetry is completely restored. 

\begin{figure*}
\centering
\includegraphics[width=7.in]{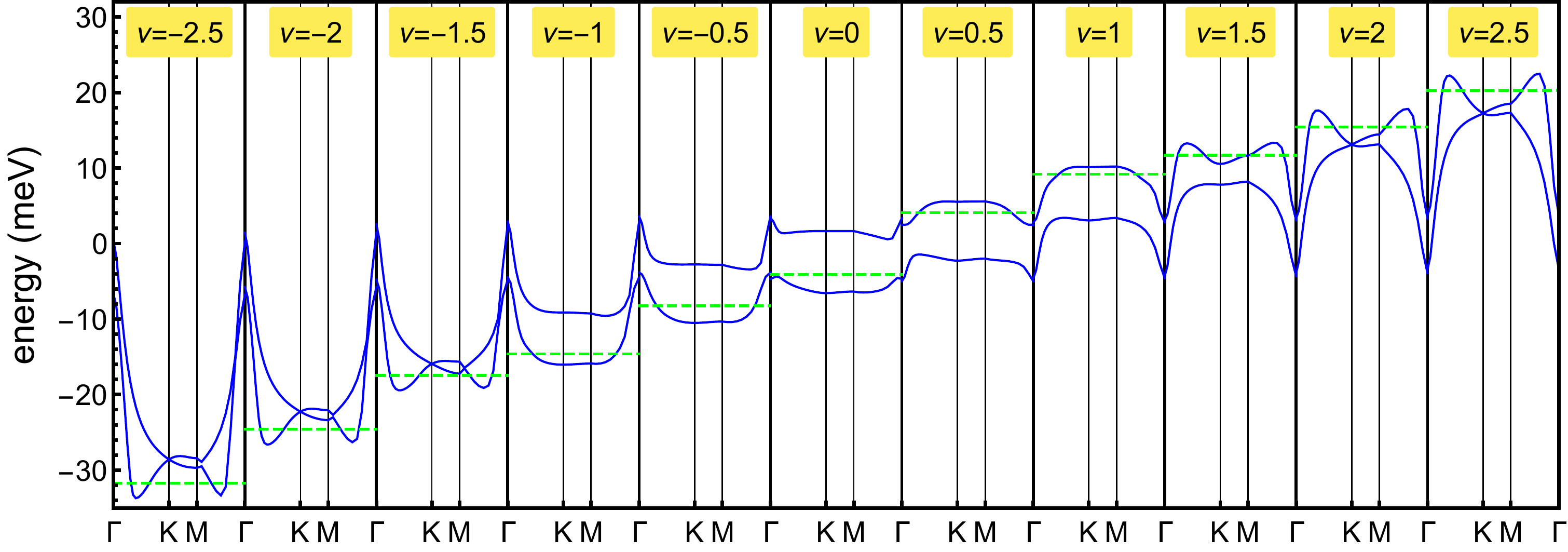}
\caption{Evolution of the band structure in the Hartree-Fock approximation upon varying the filling, for $-2.5\leq\nu\leq 2.5$. The green dashed line identifies the Fermi energy.}
\label{np_bands_evolutionSI}
\end{figure*}

The evolution of the band structure corresponding to the non-polarized solution is shown in Fig.~\ref{np_bands_evolutionSI}, for $-2.5\leq\nu\leq2.5$. As it can be seen, solutions breaking the $\mathcal{C}_2$ symmetry are not stable for $\nu>1.5$ and $\nu\le -1.5$. Two features of the band structure are worth to be further noticed: i) the lack of particle-hole symmetry, so that the bands are not symmetric upon inverting $\nu$ to $-\nu$; ii) the bands are rigid at the $\Gamma$ point of the BZ. This is a consequence of the fact that the \textcolor{black}{charge density of the TBG evaluated at the $\Gamma$ point of the Brillouin Zone is almost homogeneous as compared to the other high symmetry points of the BZ}~\cite{rademaker_prb18}, as shown in Fig.~\ref{Density}. The resulting dependence of the band shape on filling factor has been extensively discussed~\cite{Guinea_pnas18,cea_prb19,RAM19,CB20,ZJWZ20,GVLML20,XM20,LD21,LKLV21}. This significant dependence of the band shape on electron filling implies the existence of a special effective electron-electron interaction, electron assisted hopping~\cite{SV34,VK79,VK79b,HM90,MH90}.

The dependence of the Hartree-Fock potential on electron occupancy arises from the variations in charge distribution in different regions of the Brillouin Zone~\cite{rademaker_prb18}. These variations are determined by form factors of the type:
\begin{align}
    {\cal M}_{\vec{G}} ( \vec{k} , \vec{k} + \vec{q} ) &= \left\langle \vec{k} , m \left| \hat{\rho}_{\vec{q} + \vec{G}} \right| \vec{k} + \vec{q} , n \right\rangle = \int_\Omega d^2 \vec{r} u_{\vec{k} , m}^* ( \vec{r} ) e^{i \vec{G} \vec{r}} u_{\vec{k} + \vec{q} , n} ( \vec{r} )
    \label{ffactor}
\end{align}
where $\vec{G}$ is a reciprocal lattice vector, $\vec{k}$ and $\vec{k} + \vec{q}$ are in the moiré Brillouin Zone, $m$ and $n$ are band indices, $u_{\vec{k} , m} ( \vec{r} ) $ is the periodic part of the wavefunction associated to $| \vec{k} , n \rangle$, and $\Omega$ is the moiré unit cell. These from factors encode the complexity of the band wavefunctions of twisted bilayer graphene. There are some simplifying approximations made in the literature. These are very convenient because they allow for simple analytical calculations, like the "flat metric condition" used in~\cite{BSRL21,SLRB21}, however, the neglect the dependence of the form factors in Eq.~\eqref{ffactor} on the position $\vec{k}$ in the Brillouin Zone. Since the variations of the form factors over the BZ are substantial, it is not immediately clear how realistic this approximation is.

\begin{figure*}
\centering
\includegraphics[width=5.in]{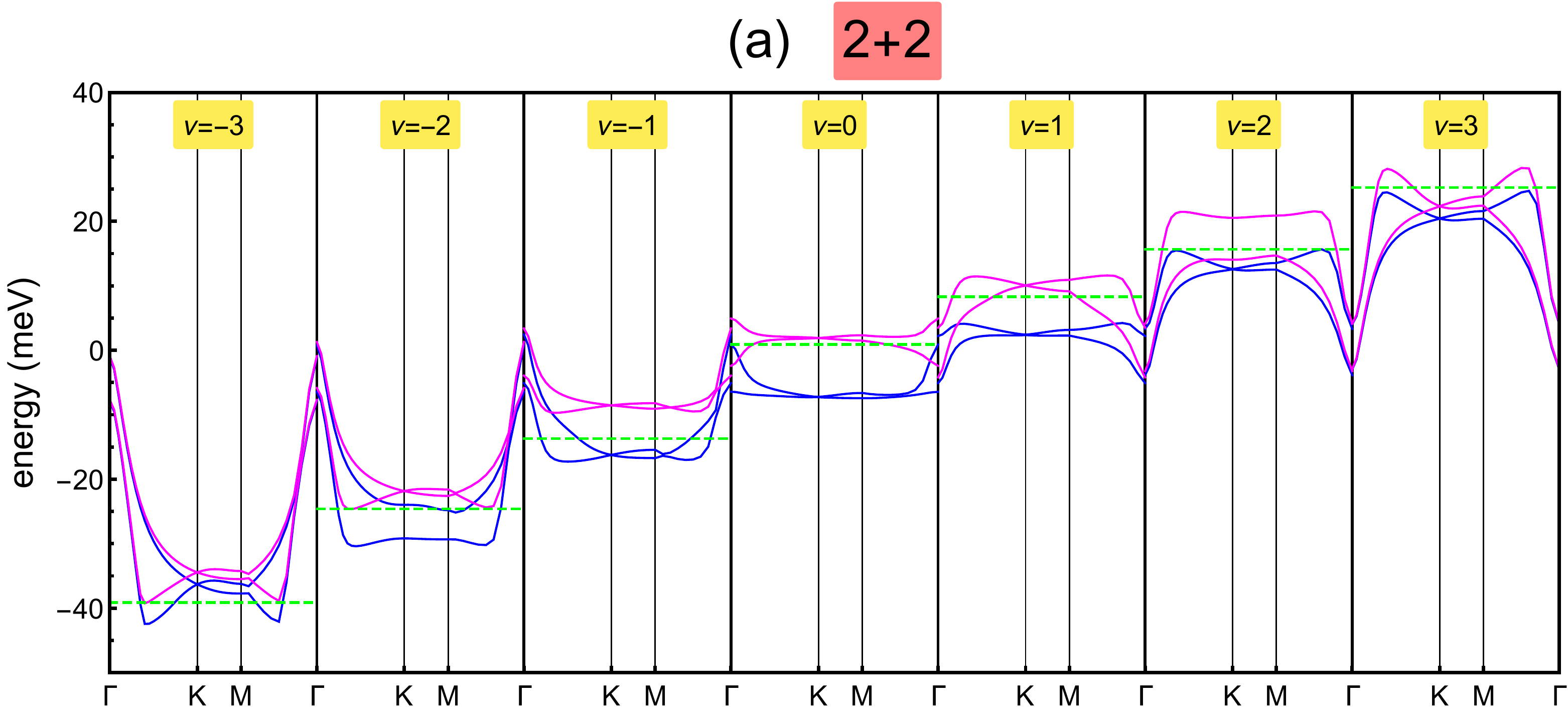}
\includegraphics[width=5.in]{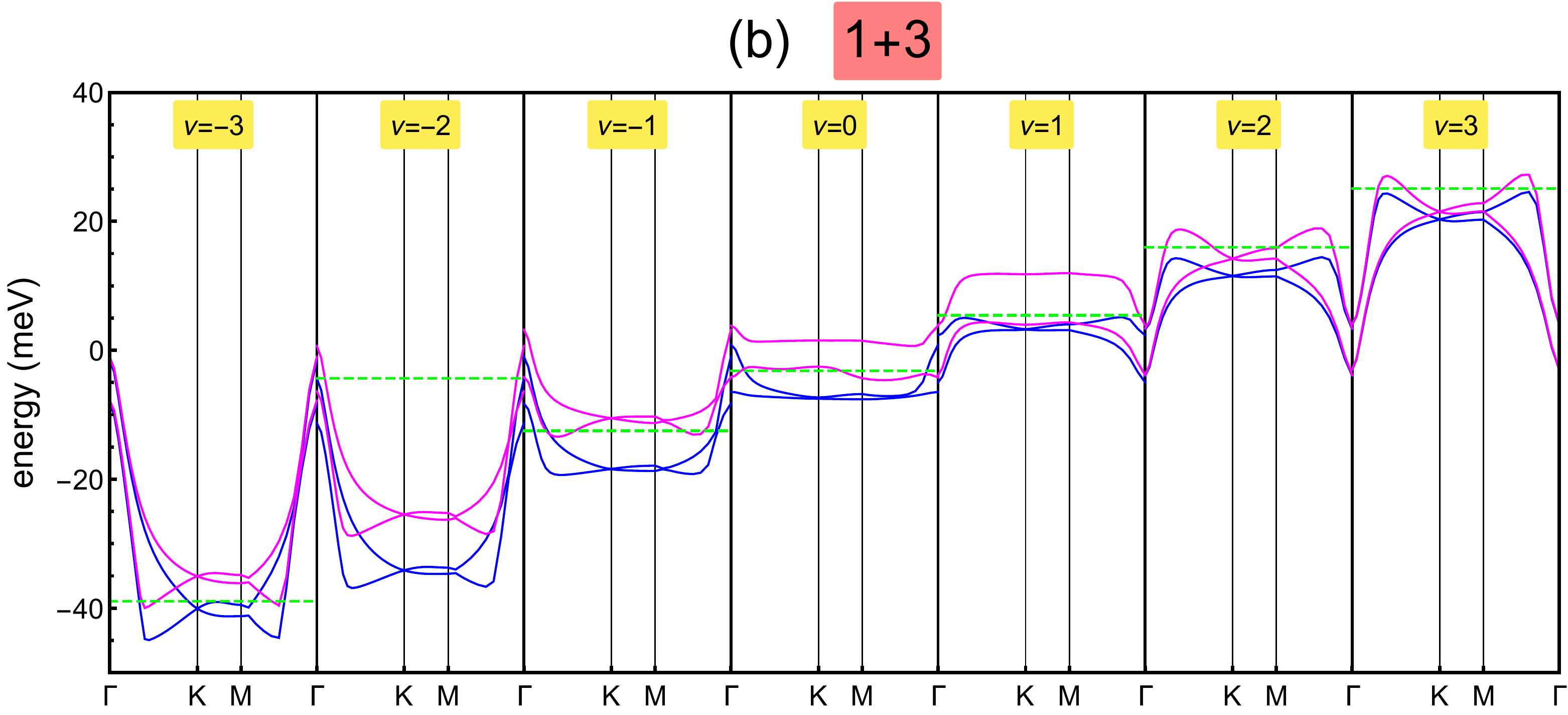}
\includegraphics[width=5.in]{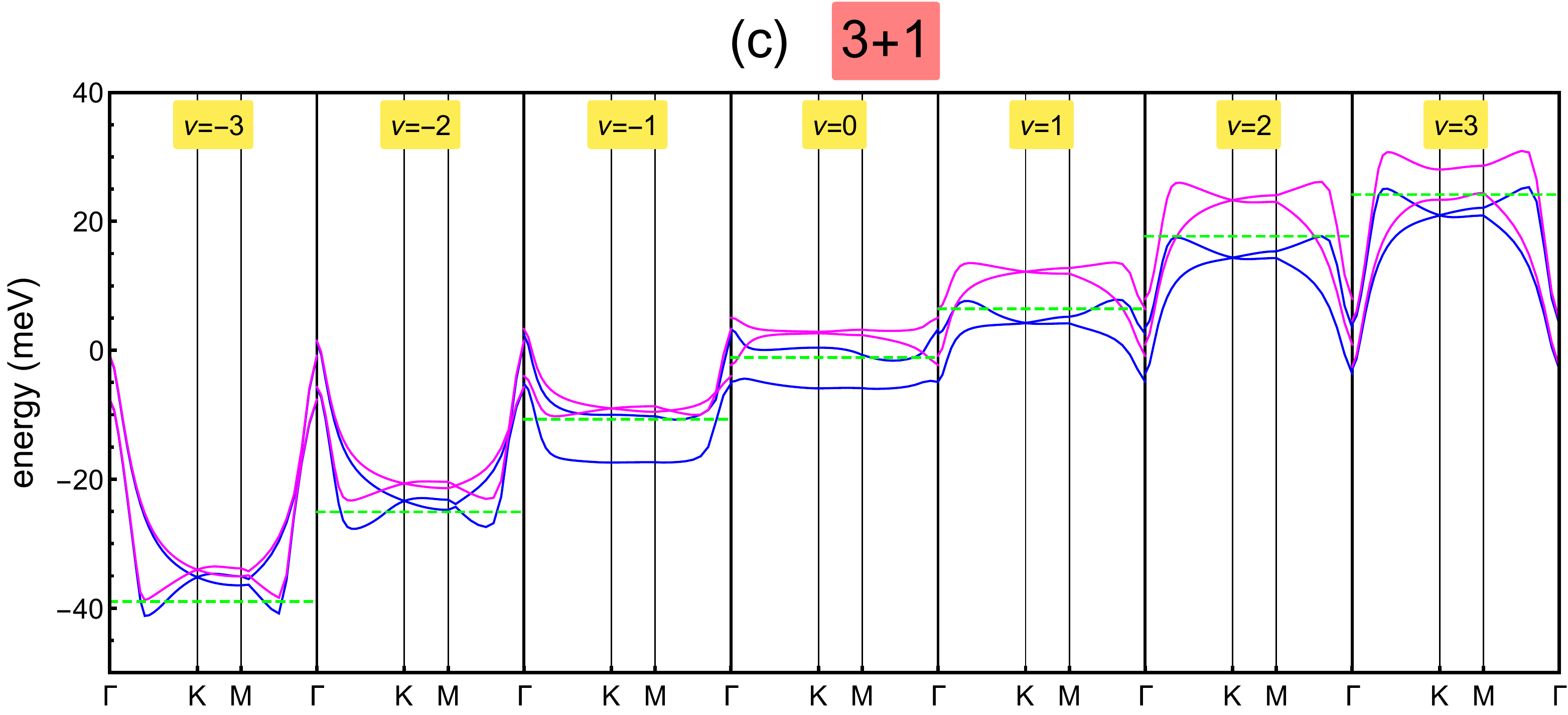}
\caption{
Band structure obtained at integer filling, $-3\le\nu\le3$,
for three different polarized solutions: $2+2$ (a), $1+3$ (b) and $3+1$ (c). The  blue and magenta lines show the bands of higher and lower occupancy, respectively. The green dashed line in each panel identifies the Fermi energy.
}
\label{polarized_bands_evolutionSI}
\end{figure*}

\begin{figure}
\centering
\includegraphics[width=4.in]{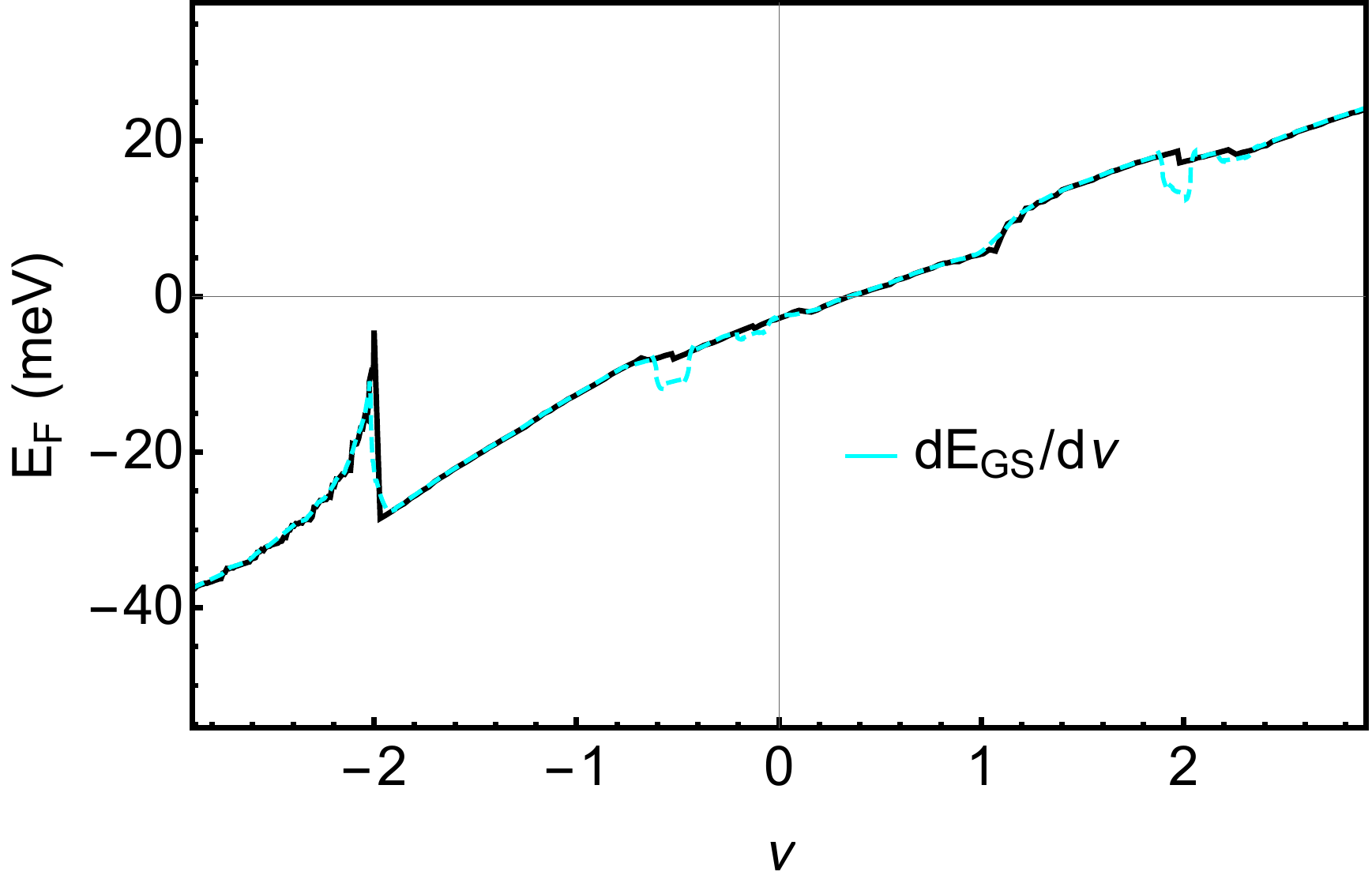}
\caption{
Fermi energy as a function of the filling for the $1+3$ solution. The continuous black line shows the value of $E_F$ as obtained from the highest energy occupied states, while the dashed cyan line represents an alternative definition of $E_F$, the derivative of the GS energy with respect to $\nu$.
}
\label{EF_vs_nu}
\end{figure}

\section{Broken symmetry solutions.}
\subsection{Polarized solutions.}
We consider first valley and/or spin polarized phases, motivated by the "cascade" of phases found experimentally~\cite{Wong2020Cascade}.
We define these phases using two integers, $n_+$ and $n_-$, so that there are $n_+$ bands with high occupancy, and $n_-$ bands with low occupancy. For a total occupancy $\nu$, such that $-4 \le \nu \le 4$, the occupancies of the two types of bands are listed in table~[\ref{tbl_occ}].

\begin{table}[htb]
\renewcommand{\arraystretch}{2}
\begin{tabular}{||c|c|c|c||}
\hline\hline
& $2+2$ &$3+1$ &$1 + 3$ 
\\ 
\hline
$\displaystyle n_-$ &$\displaystyle 4 + \frac{\nu - | \nu |}{2}$ & $\displaystyle 3 + \frac{\nu - | \nu + 2 |}{2}$&$\displaystyle 5 + \frac{\nu - | \nu - 2 |}{2}$ \\
\hline
$\displaystyle n_-$ &$\displaystyle \frac{\nu + | \nu |}{2}$ & $\displaystyle 1 + \frac{\nu + | \nu + 2 |}{2}$&$\displaystyle -1 + \frac{\nu + | \nu - 2 |}{2}$ \\
\hline \hline
    \end{tabular}
    \caption{Filling of the low occupancy bands, $n_-$, and the high occupancy bands, $n_+$, as function of the total filling, $-4 \le n_- \le 4$, see Fig.~[\ref{polarized_bands_evolutionSI}].}
    \label{tbl_occ}
\end{table}

Figure~\ref{polarized_bands_evolutionSI} shows the evolution of the band structure of the polarized solutions: $2+2$ (a), $1+3$ (b) and $3+1$ (c),
for integer fillings and $-3\le\nu\le3$. The high and low occupancy sets of bands are represented by the blue and magenta lines, respectively. Solutions breaking the $\mathcal{C}_2$ symmetry occur at $\nu=\pm2$ for the $2+2$ solution, $\nu=1$ for the $1+3$ solution and $\nu=-1,3$ for the $3+1$ solution. Interestingly, the solution $1+3$ does not break $\mathcal{C}_2$ at $\nu=-3$, in contrast to what is expected. We argue that here the $\mathcal{C}_2$ symmetry breaking is prevented by the small value of the interaction. Furthermore, it is worth noting that, in the $1+3$ solution at $\nu=-2$, the empty low occupancy bands mostly stay below the Fermi level. At this filling the Fermi energy, which we define as the highest energy of the occupied states, decreases upon increasing the filling, implying a negative compressibility. This behavior can be better seen in the Fig.~\ref{EF_vs_nu}, showing $E_F$ as a function of $\nu$ and displaying a jump-like discontinuity at $\nu=-2$. The cyan dashed line in the Fig. \ref{EF_vs_nu} represents the value of $E_F$ as obtained from the derivative of the GS energy with respect to $\nu$, which indeed matches quite well the curve of $E_F$ computed as described above. However, this anomalous behavior is actually not very meaningful in the present context, as the solution $1+3$ turns out to not be stable close to $\nu=-2$, as emphasized by Figs.~\ref{np_bands_evolutionSI} and  Fig.~\ref{polarized_bands_evolutionSI}.  

\subsection{Solutions at half filling.}

The competition between phases at half filling  has extensively been discussed~\cite{XM20,CG20,bultinck_prx20,LKLV21}. Hartree-Fock calculations at the bands near half filling have already been presented in Fig.~[\ref{bands_np}] and Fig.~[\ref{np_bands_evolutionSI}]. At half filling it is reasonable to expect that the Hartree potential vanishes (note that the continuum model allows us only to compute differences in the charge distribution at different fillings). This assumption is consistent with lattice based tight binding calculations which include the entire $p_z$ band~\cite{GVLML20}. It is worth noting that modifications of the moiré structure, such as the one induced by heterostrains, can alter significantly the effect of interactions~\cite{O20}.

We assume first that the leading mean field decoupling of the interaction is of the type 
\begin{align}
    {\cal H}_{MF}^{int} = \sum_{\vec{q}} V_{\vec{q}} \langle \hat{\rho}_{\vec{q}} \rangle \hat{\rho}_{\vec{q}} + h. c.
    \label{decoupling}
    \end{align}
    where $\hat{\rho}_{\vec{q}}$ is the density operator, eq.(\ref{densq}). Within this approximation,
 the only coupling between electrons of different flavors, valley and spin, is through the Hartree potential, as the density operator, Eq.~(\ref{densq}) is diagonal in valley space. 
 
 The lack of a Hartree potential implies that the resulting Fock hamiltonian, within the decoupling scheme in Eq.(\ref{decoupling}) is split into four independent hamiltonians, one for each spin and valley. Each of these hamiltonians admits solutions with the full symmetries of the non interacting Hamiltonian, and, at least one solution where the equivalence between the two sublattices, described by a ${\cal C}_2 {\cal T}$ symmetry is broken. The latter solution shows a gap between the valence and the conduction bands, and it has a lower energy when the valence band is fully occupied and the conduction band is empty. As in this phase the charge has a higher weight in one sublattice, we can describe the corresponding Slater determinant in terms of the spin, valley, and majority sublattice, $| Sp , V , S \rangle$, where $Sp = \uparrow , \downarrow ; V = K , K' ; S = A , B$. In each of the possible $2^3 = 8$ wavefunctions the occupied valence band is associated to a valley, so that it carries a Chern number ${\cal C} = \pm 1$, depending on the combination of valley and sublattice, $\tau_z \sigma_z$.

A possible ground state at half filling is:
\begin{align}
    | GS \rangle_\sigma &\equiv  | \uparrow , K , S_{\uparrow , K} \rangle \otimes | \downarrow , K , S_{\downarrow , K} \rangle \otimes | \uparrow , K' , S_{\uparrow , K'} \rangle \otimes | \downarrow , K' , S_{\downarrow , K'} \rangle
    \label{wv_hf}
\end{align}
where the sublattice labels $S_{Sp , V}$ can take the values $A , B$. As this sublattice index can take two values in each component, there are 16 possible wavefunctions of the type shown in Eq.~(\ref{wv_hf}). There are two wavefunctions with Chern number ${\cal C} = \pm 4$, four wavefunctions with ${\cal C} = \pm 2$, and six wavefunctions with ${\cal C} = 0$. 

Other Slater determinants which maximize the overlap between one electron wavefunctions are superpositions of states with the same valley and spin indices. These determinants are globally spin and valley polarized.
\begin{align}
    | GS \rangle_P &\equiv  | Sp_1 , V_1 , v \rangle \otimes | Sp_1 , V_1 , c \rangle \otimes | Sp_2 , V_2 , v \rangle \otimes | Sp_2 , V_2 , c \rangle
    \label{wv_pol}
\end{align}
where $Sp_1 , Sp_2$ are spin indices, $V_1 , V_2$ are valley indices, and $v , c$ stand for valence and conduction bands. In the limit when the bands are infinitely flat~\cite{SGG12,TKV19} this valley/spin polarized wavefunction is degenerate, within the Hartree-Fock approximation, with the  sublattice polarized wavefunction in Eq.~(\ref{wv_hf}).

\begin{figure}
\centering
\includegraphics[scale=0.21]{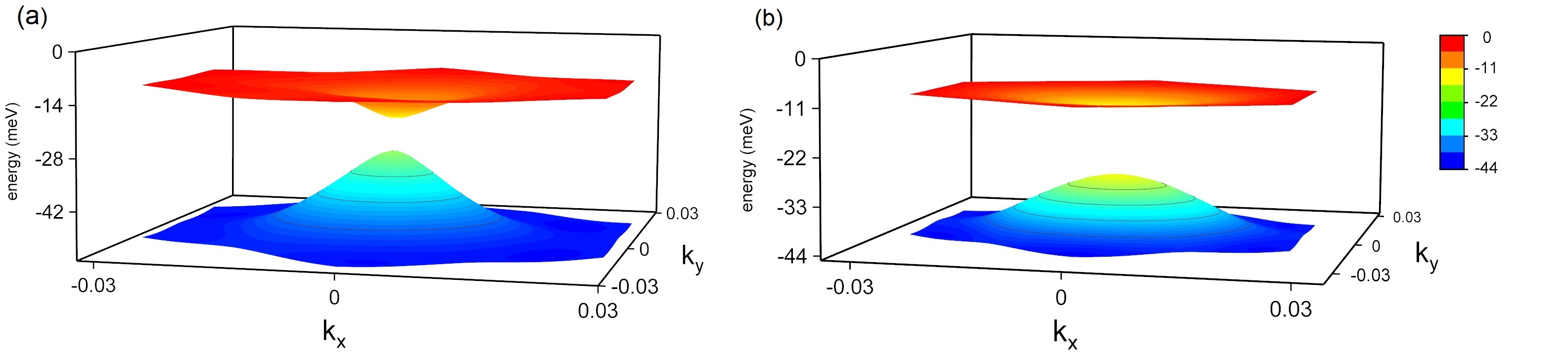}
\caption{
Hartree-Fock bands at a magic angle and at half filling, for a state of the form Eq.~(\ref{wv_hf}). We show the self-consistent solutions for initially a) dispersive bands, as in Fig.~\ref{bands_koshino}, and in b) for initially non-dispersive bands (chiral model at a magic angle)~\cite{SGG12,TKV19}. 
}
\label{bands_HF}
\end{figure}

The wavefunction in Eq.~(\ref{wv_hf}) shows four occupied valence bands, separated by a gap from four unoccupied conduction bands, (the problem reduces to four copies of the bands shown in Fig.[\ref{bands_HF}]). The valley and spin polarized wavefunction in Eq.~(\ref{wv_pol}) shows two occupied valence bands and two occupied conduction bands. In this case, there is no energy gain in opening a gap between the valence and conduction bands.We can expect, in both cases, that the overlap between one electron wavefunctions is higher between states in the same band (valence and conduction)  than between states in different bands. If that is the case, the wavefunction in Eq.~(\ref{wv_hf}) will have a lower energy, as no conduction band is occupied.

At this stage, the Hartree-Fock approximation leaves undetermined the relative phase between wavefunctions in different valleys and different sublattices. Solutions with well defined phases between Slater determinants in different valleys arise if the mean field hamiltonian allows for inter-valley matrix elements, not included in the decoupling scheme described in Eq.~(\ref{decoupling}).

%A phase between wavefunctions in different valleys will be favored if superpositions of wavefunctions like the one in Eq.~(\ref{wv_hf}) lower their energy through the emergence of inter-valley matrix elements in the exchange potential. 

Such terms can favor a inter-valley coherent (IVC) combination of the type~\cite{bultinck_prx20}:
\begin{align}
    | GS \rangle_{IVC} &\equiv \left( | \uparrow , K , A \rangle + i e^{i \phi} | \uparrow , K' , B \rangle \right) \otimes \left( | \uparrow , K , B \rangle  - i e^{-i \phi} | \uparrow , K' , A \rangle \right) \otimes \nonumber \\
    &\otimes
  \left( | \downarrow , K , A \rangle + i e^{i \phi} | \downarrow , K' , B \rangle \right) \otimes \left( | \downarrow , K , B \rangle  - i e^{-i \phi} | \downarrow , K' , A \rangle \right)
    \label{wv_IVC}
\end{align}
The existence of a gap in the one particle spectrum of the independent valley solutions shown in Fig.[\ref{bands_HF}] implies that inter-valley terms will only appear above a threshold determined by the ratio between the inter-valley interactions and the gap. In order to check for  the emergence of inter-valley terms in the Hartree-Fock hamiltonian, we have carried out calculations allowing for off-diagonal valley terms in the Hartree-Fock hamiltonian, projecting the hamiltonian onto the two central bands. The effect of intervalley biases in the initial wavefunctions tends to vanish in the iteraction towards self consistency, and the bands with and without initial intervalley bias agree within numerical accuracy. These results suggest that the valleys remain decoupled.
 
 Interactions at the atomic scale, will favor a phase in which the two sublattices have equal occupation, over the phase where all the charge resides in the same sublattice~\cite{AF06,GLGS17,SRRN19,KGMKL21}. Finally, the zeroth point energy of excitations above the Hartree-Fock solutions discussed here can also favor specific combinations of wavefunctions of the type shown in Eq.~(\ref{wv_hf}) with different Chern numbers~\cite{KCBZV21}, see Eq.~(\ref{wv_IVC}).

%The description of the long range interactions in the continuum model relies on matrix elements of the density operator, Eq.~(\ref{densq}), which is diagonal in spin, valley, sublattice, and layer spaces. Hence, the Hartree-Fock solutions mentioned here are invariant under rotations in spin and valley spaces. It has been argued that the band dispersion, and Hund like interactions may favor a special type of ground state~\cite{bultinck_prx20}, the intervalley coherent state (IVC), which can be of various types. A possible one is characterized by a finite value of the operator $\sigma_y \tau_+ e^{i \phi} + {\rm h. c.}$, where the operators $\sigma$ and $\tau$ act on the sublattice and valley degrees of freedom, and $\phi$ is a phase. Using the same notation as in Eq.~(\ref{wv_hf}), a possible state of this type can be written as:

%This wave function is a superposition of states of the type shown in Eq.~(\ref{wv_hf}). These are degenerate in the Hartree-Fock approximation, which is thus insufficient to resolve the difference in energy between the state in Eq.~(\ref{wv_IVC}) and  states derived from other combinations of wavefunctions of the type in Eq.~(\ref{wv_hf}). These energy difference are likely to be very small.

It is worth noting that, even in the absence of the Hartree potential, the exchange term induces significant deformations in the band dispersion. The electronic bands associated to the state in Eq.~(\ref{wv_hf}) for initially dispersive and for the non dispersive bands resulting from the chiral model~\cite{SGG12,TKV19} are shown in Fig.~\ref{bands_HF}.

\section{Conclusions.}
We have discussed the electronic properties of interacting electrons in twisted bilayer graphene obtained using the mean field Hartree-Fock approximation. This is usually the initial method to study interactions in theoretical condensed matter physics, and it is also a reasonable starting point to analyze effects beyond mean field theory. The Hartree-Fock method has been applied to the study of the long range electrostatic interactions. Simple scaling arguments suggest that this is the leading interaction between electrons in twisted bilayer graphene. 

Despite the simplicity of the method, the Hartree-Fock analysis of interactions in twisted bilayer graphene shows a number of features not commonly found in most materials studied in condensed matter physics. Among them, we can emphasize:
\begin{itemize}
    \item 
    A significant distortion of the band shapes as function of the electron filling. This effect is mostly driven by the Hartree term, although it is also significant at half filling, solely due to the exchange potential.
    
    This filling dependence makes twisted bilayer graphene a system where a good case can be made for the existence of electron assisted hopping.
    \item 
    The existence of a gapped phase at half filling, driven by the exchange term. 
    
    This phase is sublattice polarized in each flavor (spin and valley) sector. The Hartree-Fock  allows for many combinations of these separate sectors.
    \item
    Away from half filling, the Hartree-Fock approximation leads to spin and/or valley polarized phases. As at half filling, the Hartree-Fock approximation respects the spin and valley degeneracy, so that the details of the polarization cannot be resolved.
    \item
    The energy differences between the polarized phases away from half filling are small, of the order of a few meV per unit cell. At these scales, other interactions, like the intra-atomic Hubbard repulsion, are likely to play a role, and select the most stable ground state.
    
\end{itemize}

\section*{Acknowledgements}
This work was supported by funding from the European Commission, under the Graphene Flagship, Core 3, grant number 881603, and by the grants NMAT2D (Comunidad de Madrid, Spain),  SprQuMat and SEV-2016-0686, (Ministerio de Ciencia e Innovación, Spain). NRW is supported by STFC grant ST/P004423/1. We are thankful to H. Ochoa ant to M. A. Cazalilla for helpful discussions, and to M. I. Katsnelson for pointing out to the references~\cite{SV34,VK79,VK79b}. 

\bibliography{Literature}
\bibliographystyle{apsrev4-1}
\end{document}